# Thermodynamic properties of the solid and metal electrons in the nonextensive quantum statistics with a nonextensive parameter transformation

Yahui Zheng[1], Jiulin Du[2]
1: *School of Science, Henan Institute of Technology, Xinxiang City 453003, China.*
2: *Department of Physics, School of Science, Tianjin University, Tianjin 300072, China.*

**Abstract** We study the thermodynamic properties of solid and metal electrons in the nonextensive quantum statistics with a nonextensive parameter transformation. First we study the nonextensive grand canonical distribution function and the nonextensive quantum statistics with a parameter transformation. Then we derive the generalized Boson distribution and Fermi distribution in the nonextensive quantum statistics. Further we study the thermodynamic properties of solid and metal electrons in the nonextensive quantum system, including the generalized Debye models, the generalized internal energies, the generalized capacities and chemical potential. We derive new expressions of these thermodynamic quantities, and we show that they all depend significantly on the nonextensive parameter and in the limit they recover to the forms in the classical quantum statistics. These new expressions may be applied to study the new characteristics in some nonextensive quantum systems where the long-range interactions and/or long-range correlations play a role.

## 1 Introduction

In recent years, a lot of attention has been paid to nonextenstive statistical mechanics (NSM), generated from the work of Tsallis [1], where one new entropy with nonextensive property was proposed. The main feature of NSM is that it always leads to one type of power-law distribution functions. Therefore, it is regarded as a generalization of Boltzmann-Gibbs (BG) statistical theory, so as to deal with some complex systems with non-exponential or power-law distributions. NSM has been proved to be a very effective theory for the nonequilibrium complex systems in external fields [2-9] and, therefore, has been applied to many interesting fields, such as astrophysical systems [2-5], plasma physics [6-9], anomalous diffusion systems [10,11], biological and chemical systems [12-14] and so on.

In NSM, the deduction of the power-law distribution can be carried out through two primary paths. The first is through the maximum entropy procedure, with some constraints. In this aspect, there were at least four methods to get the power-law distribution, such as the original method [1], the un-normalized method [15], the normalized method [16], and the optimal Lagrange multiplier method [17]. The second is through treatment of statistical ensemble, in which several approaches are also developed, including the dynamical thermostatting approach [18], the counting rule [19], the fractal dimension of phase space [20], and the order parameter into the generalized Lorentzian [21].

Now, our main concern is the nonextensive quantum statistics, whose base is the grand canonical distribution, derived from the equiprobability principle of statistical ensemble. In order to develop an appropriate formalism of nonextensive quantum statistics, there have been at least two prescriptions to be proposed.

The first one is the factorization approximation [22], which is used to deal with the quantum sum of the grand partition function. From this prescription, simple statistical expressions for the quantum systems can be obtained, and therefore it is widely adopted in many applications such as the black body radiation [23-24], early universe [25-26], the Bose-Einstein condensation [27-29], the superconductivity [30-31], and the spin systems [32-35]. But the disadvantage of this method is also very obvious that the statistical expression obtained with the factorization approximation is merely applicable in the case $\beta n_i \to 0$, where the $\beta$ is the usual inverse temperature and $n_i$ is the occupation number in $i$th microstate for given quantum system. Apparently, it further corresponds to two special cases, namely, the case with very high temperature and the case with very small occupation number, which indicates the considered quantum system is very dilute, called dilute gas approximation [22-24]. On the other hand, the practical applications of this prescription have already exceeded the case $\beta n_i \to 0$.

On the contrary, the second one is accurate in math, where Green function is introduced to the formally integral representation for the grand canonical partition function [36], as the classical Boltzmann-Gibbs counterpart. It can be seen that due to the integral representation by recourse to Green function, complication in math is striking. This directly leads to the difficulty of applying it to realistic systems. Instead of this, people must adopt some approximations for the formally integral representation as to proceed in the integral calculations for some real systems [36-37].

In the present work, we study the thermodynamic properties of solid and metal electrons in the nonextensive quantum statistics with an approach called parameter transformation. In section 2, we study the grand canonical distribution function on the basis of the equal probability principle. In section 3, we study the nonextensive quantum statistics with a parameter transformation. In sections 4 and 5 respectively, we apply the nonextensive quantum statistics to study the thermodynamic properties of solid and metal electrons. And in section 6, we give the conclusions.

## 2 The nonextensive grand canonical distribution based on the equal probability principle

In the grand canonical ensemble, the system is on contact with a huge reservoir, thus permitting both the particle and energy exchanges between the system and the reservoir. Let us combine the system with the reservoir as one composite isolated system, whose entropy, according to the equal probability principle, is given by the $q$-logarithm nonextensive form [38], namely,

$$S_q = k \ln_q \Omega \equiv k \frac{\Omega^{1-q} - 1}{1 - q}, \tag{1}$$

where the quantity $\Omega$ is the number of the quantum state of the composite isolated system. And the Tsallis factor is defined as

$$c_q \equiv 1 + (1-q)\frac{S_q}{k} = \Omega^{1-q}. \tag{2}$$

In an ordinary nonextensive complex system, energies of the system at different quantum states are essentially nonadditive, namely,

$$E_{ij} = E_i + E_j - (1-q)\beta' E_i E_j, \tag{3}$$

where $\beta'$ is the generalized Lagrange multiplier [39], "$i$" and "$j$" represent different quantum states. The cross term in (3) is derived from the interactions between different parts or states inside the system. In the following discussions, we would show that the value of (1-$q$) is very tiny. So we can neglect the cross term in (3) and adopt the additive approximation. In view of this, setting the total particle number of this composite system as $N_0$ and the total energy as $E_0$, we have

$$E_0 \simeq E_s + E_r, \quad N_0 = N + N_r, \tag{4}$$

where $E_r$ and $N_r$ are the reservoir's energy and particle number respectively, $E_s$ and $N$ are energy and particle number respectively of the system at the quantum state $s$.

We assume the reservoir is very huge, then there should be

$$E_s \ll E_0, \quad N \ll N_0. \tag{5}$$

According to the equiprobability principle, the probability distribution of the system with energy $E_s$ and particle number $N$ is proportional to the quantum state number of the reservoir with energy $E_r$ and particle number $N_r$, that is to say,

$$\rho_{Ns} \sim \Omega_r(E_0 - E_s, N_0 - N). \tag{6}$$

Considering the form of the entropy (1), let us consider the Taylor expansion of the quantity $\ln_q \Omega_r$, and only retain the first three terms. Then we have

$$\ln_q \Omega_r(N_0 - N, E_0 - E_s)$$
$$= \ln_q \Omega_r(N_0, E_0) + \left(\frac{\partial \ln_q \Omega_r}{\partial N_r}\right)_{N_r = N_0} (-N) + \left(\frac{\partial \ln_q \Omega_r}{\partial E_r}\right)_{E_r = E_0} (-E_s)$$
$$= \ln_q \Omega_r(N_0, E_0) - \alpha_r N - \beta_r E_s, \tag{7}$$

where two Lagrange multipliers are introduced that

$$\alpha_r = \left(\frac{\partial \ln_q \Omega_r}{\partial N_r}\right)_{N_r = N_0}, \quad \beta_r = \left(\frac{\partial \ln_q \Omega_r}{\partial E_r}\right)_{E_r = E_0}, \tag{8}$$

and then we get

$$\rho_{Ns} \sim [1 - (1-q)(\alpha_r N + \beta_r E_s)/c_{qr}]^{\frac{1}{1-q}}. \tag{9}$$

Considering the balance conditions [39], we introduce the following generalized multipliers,

$$\frac{\alpha_r}{c_{qr}} = \frac{\alpha}{c_q} \equiv \alpha', \quad \frac{\beta_r}{c_{qr}} = \frac{\beta}{c_q} \equiv \beta'. \tag{10}$$

Here, $c_{qr}$ is the Tsallis factor of the reservoir.

Therefore, the grand ensemble distribution function is given by

$$\rho_{Ns} = \frac{1}{\Xi_q}[1 - (1-q)(\alpha' N + \beta' E_s)]^{\frac{1}{1-q}}, \tag{11}$$

where the grand partition function can be defined as

$$\Xi_q = \sum_{N=0}^{\infty} \sum_s [1 - (1-q)(\alpha' N + \beta' E_s)]^{\frac{1}{1-q}}. \tag{12}$$

This grand canonical distribution function is what we need in next sections. The Tsallis factor now is given by

$$c_q = \sum_{N=0}^{\infty} \sum_s \rho_{Ns}^q. \tag{13}$$

This expression is not the same as the original Tsallis factor in (2). Express (13) holds in the open system in (statistical) equilibrium while (2) is valid for an isolated system at (statistical) equilibrium state. Lastly, let us give the definitions of q-average of the particle number and internal energy of the system by

$$\bar{N} = \frac{\sum_{N=0}^{\infty}\sum_{s} N \rho_{Ns}^{q}}{\sum_{N=0}^{\infty}\sum_{s} \rho_{Ns}^{q}}, \quad U = \frac{\sum_{N=0}^{\infty}\sum_{s} E_{s} \rho_{Ns}^{q}}{\sum_{N=0}^{\infty}\sum_{s} \rho_{Ns}^{q}}. \quad (14)$$

This definition approach is of the third choice of internal energy definition [40], from which one can find that the particle number and internal energy are both additive. This energy is called Lagrange internal energy [39,41]. Similarly, the particle number is called the Lagrange particle number.

**3 The nonextensive quantum statistics with a parameter transformation**

We consider the nonextensive quantum system, where the long-range interactions or/and correlations lead to the correlations of quantum states of the system. The quantum states can be regarded as the superposition of single-particle quantum states, marked by the occupation number. In order to deduce the expressions for the averaged occupation number in nonextensive statistics, let us consider a system consisting of $N$ identical particles, with the distribution of occupation numbers described by $\{a_n, n=1,2,3...\}$, where $n$ denotes the single-particle quantum state of the system. The energy spectrum is given by $\{\varepsilon_n, n=1,2,3...\}$.

Then the particle number and energy are expressed as

$$N = \sum_{n} a_n, \quad E_s = \sum_{n} \varepsilon_n a_n, \quad (15)$$

In the grand canonical ensemble, the distribution $\{a_n\}$ is indefinite, so we must consider all the possibilities in the grand canonical sum. According to the definitions (15) and (14), the averaged occupation number at given quantum state of the energy is defined as

$$\bar{a}_n = \frac{\sum_{N=0}^{\infty}\sum_{s} a_n \rho_{Ns}^{q}}{\sum_{N=0}^{\infty}\sum_{s} \rho_{Ns}^{q}} = \frac{\sum_{\{a_l\}} a_n [1-(1-q)\sum_{l}(\alpha'+\beta'\varepsilon_l)a_l]^{\frac{q}{1-q}}}{\sum_{\{a_l\}}[1-(1-q)\sum_{l}(\alpha'+\beta'\varepsilon_l)a_l]^{\frac{q}{1-q}}}. \quad (16)$$

In order to evaluate this averaged occupation number, we need to factorize the expression of the above quantum sum. The direct factorization approximation demands the system temperature must be very high. To avoid this, the following parameter transformation is introduced,

$$(1-q)\sum_{n}(\alpha'+\beta'\varepsilon_n)a_n = 1-\nu. \quad (17)$$

On the left side of the above transformation, the summation term is proportional to the total freedom degree of whole system, in a given distribution of occupation number. The main intention of the transformation (17) is to replace the nonextensive parameter $q$ with the new nonextensive parameter $\nu$. Here, we assume the whole system to share the same $\nu$, which means that the value of $q$ changes with the variation of the distribution of occupation number. But in the subsystem with fixed distribution of occupation number, the value of $q$ is fixed. Further if averaging (17) over all possible distributions of occupation numbers, there is a relation between two nonextensive parameters $q$ and $\nu$:

$$(1-q)N\frac{d}{2} = 1-\nu, \quad (18)$$

where $N$ is the (averaged) particle number of the system and $d$ is the freedom degree of single particle. Due to the tiny value of (1-*q*), which would be shown later on, we have supposed that the system share the same *q* in the averaging process.

Now let us give a short comment about the physical meanings of the above relation (18). Firstly, it should be noted that this treatment actually constructs the link between two different statistical mechanical formalisms. The first one is the statistical ensemble, whose probability distribution is a function of energy. The second one is the quasi-independent particle system, which is described by the single-particle distribution function.

Secondly, the relation (18) endows the two formalisms with two different nonextensive parameters. That is to say, these parameters in (17) and (18) are both realistic in the physical applications. This is why in many discussions about the parameter (1-*q*) for these systems with ensemble distribution, the parameter is always related to the particle number, or limited by it [42]. On the other hand, the value of (1-*v*) is irrelevant to the particle number. For instance, in the sun by use of some data of standard models, the value of (1-*v*) in the nonextensive molecular dynamics is evaluated as about 0.2 [43-44]. In ordinary cases, the value of (1-*v*) is much less than 0.2 (but still more than zero) [45]; this means that the value of (1-*q*) is very tiny. So the additive approximation of energy in (4) is reasonable.

For convenience, the Nature Index can be generalized by the following $e_g$:

$$e_g = v^{\frac{1}{v-1}}, \quad [e_g]_{v=1} = e. \tag{19}$$

Then, in view of the transformation (17), the averaged occupation number (16) becomes

$$\bar{a}_n = \frac{\sum_{\{a_l\}} a_n e_g^{-q\sum(\alpha'+\beta'\varepsilon_l)a_l}}{\sum_{\{a_l\}} e_g^{-q\sum(\alpha'+\beta'\varepsilon_l)a_l}} = \frac{\sum_{\{a_l\}} a_n e_g^{-\sum(\alpha'+\beta'\varepsilon_l)a_l+1-v}}{\sum_{\{a_l\}} e_g^{-\sum(\alpha'+\beta'\varepsilon_l)a_l+1-v}}$$

$$= \frac{\sum_{a_n} a_n e_g^{-(\alpha'+\beta'\varepsilon_n)a_n} \prod_{m \neq n} \sum_{a_m} e_g^{-(\alpha'+\beta'\varepsilon_m)a_m}}{\sum_{a_n} e_g^{-(\alpha'+\beta'\varepsilon_n)a_n} \prod_{m \neq n} \sum_{a_m} e_g^{-(\alpha'+\beta'\varepsilon_m)a_m}}$$

$$= \frac{\sum_{a_n} a_n e_g^{-(\alpha'+\beta'\varepsilon_n)a_n}}{\sum_{a_n} e_g^{-(\alpha'+\beta'\varepsilon_n)a_n}}. \tag{20}$$

In order to continue, we further define the following quantity *t*,

$$t \equiv e_g^{-(\alpha'+\beta'\varepsilon_n)} = v^{\frac{(\alpha'+\beta'\varepsilon_n)}{1-v}}. \tag{21}$$

It is obvious that this quantity has the value range 0< *t* <1, no matter what the value of (1-*v*) is. Therefore, using (20) and (21), for the Fermions in the nonextensive quantum statistics, we have the generalized Fermi distribution,

$$\bar{a}_{nF} = \frac{t}{t+1} = \frac{1}{e_g^{\alpha'+\beta'\varepsilon_n}+1}, \tag{22}$$

and, for the Bosons in the nonextensive quantum statistics, we have the generalized Boson distribution,

$$\bar{a}_{nB} = \frac{t\sum_{a_n=0}^{\infty} a_n t^{a_n-1}}{\sum_{a_n=0}^{\infty} t^{a_n}} = (1-t)t\frac{\partial}{\partial t}\sum_{a_n=0}^{\infty} t^{a_n} = \frac{1}{e_g^{\alpha'+\beta'\varepsilon_n}-1}. \qquad (23)$$

It is interesting that the above two new formulae (22) and (23) both have very concise mathematical forms and look very familiar to us. In the deduction of the two formulae, we do not adopt any mathematical approximation, indicating that the parameter transformation is an exact prescription to deal with the factorization problem in nonextensive quantum statistics.

**4 Property of solid in the nonextensive quantum statistics**

4.1 *The Lagrange internal energy and the generalized Debye model*

Just as the normal one, it is considered that the nonextensive solid system consists of $N$ particles with strong interactions, in which case the long-range correlations of particles appear. However, by use of the ordinary linear transformation of canonical coordinates, this system can be converted into that consisting of $3N$ approximately independent oscillators. According to the quantum mechanics, the energy of every oscillator is quantized. The system here can be regarded as a quantum system.

Therefore, the total energy of the system at the quantum state $s$ can be written [46] as

$$E = \varphi_0 + \sum_{i=1}^{3N} \hbar\omega_i(n_i+\tfrac{1}{2}), \quad n_i = 0,1,2,... \qquad (24)$$

where $\varphi_0$ is the zero-order term of a Taylor expansion of the system potential energy, which should be negative; $\omega_i$ is the angular frequency of the $i$th oscillator, and $n_i$ is its quantum number. In such a nonextensive system, the grand canonical distribution is given by

$$\rho_s = \frac{1}{\Xi_q}[1-(1-q)\beta' E_s]^{\frac{1}{1-q}}, \qquad (25)$$

where the subscript $s$ represents the quantum state of the whole system, and the grand canonical partition function is defined as

$$\Xi_q = \sum_s [1-(1-q)\beta' E_s]^{\frac{1}{1-q}}. \qquad (26)$$

Here we should emphasize that in the above expression (26) the quantum summation and the oscillation model summation are both considered. Therefore, the function (25) still belongs to the grand canonical distribution. In view of this, according to Eq.(14) the Lagrange internal energy of the nonexensive solid system is then

$$U = \frac{\sum_s E\rho_s^q}{\sum_s \rho_s^q} = U_0 + \sum_i \bar{n}_i \hbar\omega_i, \qquad (27)$$

where the quantity

$$U_0 = \varphi_0 + \sum_i \tfrac{1}{2}\hbar\omega_i \qquad (28)$$

should be negative, and the averaged quantum number in Eq.(27) is given by

$$\bar{n}_i = \frac{\sum_s n_i [1-(1-q)\beta' E_s]^{\frac{q}{1-q}}}{\sum_s [1-(1-q)\beta' E_s]^{\frac{q}{1-q}}}. \tag{29}$$

Similarly, we put forward the following parameter transformation,

$$(1-q)\beta' E_s = 1-v. \tag{30}$$

Again, the left side of this transformation is related to the total freedom degree of the system in the given distribution of quantum number. The average of (30) over all possible distributions of quantum numbers gives out the same result as that in (18), where for the nonextensive solid system there is $d=6$. We can also define the generalized *Nature Index*, for $v>0$,

$$e_g = v^{\frac{1}{v-1}}, \quad e_g\big|_{v=1} = e. \tag{31}$$

Then according to (20), Eq. (29) becomes

$$\bar{n}_i = \frac{\sum_{\{n_j\}} n_i e_g^{-\beta' \sum_j n_j \hbar \omega_j}}{\sum_{\{n_j\}} e_g^{-\beta' \sum_j n_j \hbar \omega_j}} = \frac{\sum_{n_i=0}^{\infty} n_i e_g^{-\beta' n_i \hbar \omega_i}}{\sum_{n_i=0}^{\infty} e_g^{-\beta' n_i \hbar \omega_i}} = \frac{1}{e_g^{\beta' \hbar \omega_i} - 1}. \tag{32}$$

Therefore, the (Lagrange) internal energy of the nonextensive solid system is written as

$$U = U_0 + \sum_i \frac{\hbar \omega_i}{e_g^{\beta' \hbar \omega_i} - 1}. \tag{33}$$

Now we employ Debye model, which regards the solid as continuous elastic medium and where there are two kinds of different waves, that is, the longitudinal wave with speed $c_l$ and the transverse wave with speed $c_t$. Then the frequency spectrum [46] reads

$$g(\omega)d\omega = B\omega^2 d\omega, \tag{34}$$

where $g(\omega)$ is the spectrum density, and the proportionality coefficient $B$ is given by

$$B = \frac{V}{2\pi^2}\left(\frac{1}{c_l^3} + \frac{2}{c_t^3}\right). \tag{35}$$

Here $V$ is the volume of the nonextensive solid system as continuous medium.

Now that the total oscillator number of the system is $3N$, there must be a maximum angular frequency $\omega_D$, satisfying [46],

$$\int_0^{\omega_D} B\omega^2 d\omega = 3N, \quad \omega_D^3 = \frac{9N}{B}. \tag{36}$$

In nonextensive statistics, we can introduce the definition of temperature duality [40], namely, there can be two temperature explanations:

$$T_q = \frac{1}{k\beta'}, \quad T = \frac{1}{k\beta}, \tag{37}$$

where $T_q$ is the physical temperature and $T$ is the Lagrange temperature, which are linked through

Tsallis factor,

$$T_q = c_q T. \tag{38}$$

Then the (Lagrange) internal energy now becomes

$$U = U_0 + B \int_0^{\omega_D} \frac{\hbar \omega^3}{e_g^{\hbar\omega/kT_q} - 1} d\omega. \tag{39}$$

If we use the following parameters,

$$y = \frac{\hbar\omega}{kT_q}, \quad x = \frac{\hbar\omega_D}{kT_q} = \frac{\theta_D}{T_q}, \tag{40}$$

where $\theta_D$ is the Debye characteristic temperature, determined by the experiments, and further we introduce the generalized Debye function,

$$D(x) = \frac{3}{x^3} \int_0^x \frac{y^3}{e_g^y - 1} dy, \tag{41}$$

and then we have

$$U = U_0 + 3NkT_q D(x), \tag{42}$$

which both depend on the nonextensive parameter $\nu$. It is clear that when we take the parameter $\nu \to 1$, Eqs.(41) and (42) recover to the forms in the classical quantum statistics[46]

4.2 *The generalized Debye model in the limit of high temperature*

Now we consider the Debye model of Lagrange internal energy of the solid in the nonextensive quantum statistics. In the limit of high temperature, the $x$ in (40) is much less than unity, so the generalized Debye model Eq. (41) becomes

$$D(x) \approx \frac{3}{x^3 \ln e_g} \int_0^x y^2 dy = \frac{1}{\ln e_g}. \tag{43}$$

Then according to (42) we obtain the internal energy:

$$U = U_0 + \frac{3NkTc_q}{\ln e_g}. \tag{44}$$

It is clear that the internal energy depends on the Lagrange temperature. In order to verify this, we need to evaluate the Tsallis factor. For this aim, we introduce the relation [40] in nonextensive statistics,

$$c_q = \Xi_q^{1-q}, \tag{45}$$

where, according to Eqs.(24), (26) and (30), the grand partition function can be evaluated as

$$\Xi_q = \sum_{\{n_i\}} e_g^{-\beta U_0 - \sum_i n_i \beta \hbar \omega_i} = e_g^{-\beta U_0} \prod_i \sum_{n_i=0}^\infty e_g^{-n_i \beta \hbar \omega_i} = e_g^{-\beta U_0} \prod_i \frac{1}{1 - e_g^{-\beta \hbar \omega_i}}, \tag{46}$$

where, for simplicity, the generalized Lagrange multiplier $\beta'$ has been replaced by the Lagrange one $\beta$. For the sake of calculation, from Eq.(46), we have

$$\ln \Xi_q = -\beta U_0 \ln e_g - \sum_i \ln\left(1 - e_g^{-\beta \hbar \omega_i}\right) \approx \sum_i \ln\left(\frac{1}{\ln e_g}\frac{kT}{\hbar \omega_i}\right), \qquad (47)$$

where the last step employs the high temperature approximation. Using the Debye model, Eq. (47) is changed as

$$\ln \Xi_q = \int_0^{\omega_D} \ln\left(\frac{1}{\ln e_g}\frac{kT}{\hbar \omega}\right) B\omega^2 d\omega = 3N \ln\left(\frac{kTe^{\frac{1}{3}}}{\ln e_g \hbar \omega_D}\right), \qquad (48)$$

According to the equality (18), in nonextensive solid system there is the relation

$$(1-q)3N = 1-\nu. \qquad (49)$$

Substituting Eq.(48) into Eq.(45), and using Eq. (44) we get the generalized internal energy,

$$U = U_0 + \frac{3NkT}{(\ln e_g)^{2-\nu}}\left(\frac{Te^{\frac{1}{3}}}{\theta_D}\right)^{1-\nu}, \qquad (50)$$

and the generalized heat capacity

$$C_V = \frac{3Nk}{(\ln e_g)^{2-\nu}}\left(\frac{Te^{\frac{1}{3}}}{\theta_D}\right)^{1-\nu}(2-\nu). \qquad (51)$$

It is interesting that the heat capacity (51) of the solid now depends on some power of the Lagrange temperature in the nonextensive quantum statistics. It is clearly that Eq.(50) and Eq.(51) both depend significantly on the nonextensive parameter ν and when we take ν→1, they recover to the forms in the classical quantum statistics.

4.3 *The generalized Debye model in the limit of low temperature*

Next, let us consider the generalized Debye model in the case of low temperature. In the limit of low temperature, the quantity $x$ in (40) is very large. Therefore the integral upper limit in Eq.(41) may be infinite. Then we have that

$$D(x) = \frac{3}{x^3}\int_0^\infty \frac{y^3}{e_g^{\,y}-1}dy = \frac{3}{x^3(\ln e_g)^4}\sum_{m=1}^\infty \frac{1}{m^4}\int_0^\infty y^3 e^{-y}dy = \frac{\pi^4}{5x^3(\ln e_g)^4}. \qquad (52)$$

Then from (42) one gets

$$U = U_0 + 3Nk\frac{\pi^4}{5(\ln e_g)^4}\frac{(Tc_q)^4}{\theta_D^3}. \qquad (53)$$

In this case, the grand partition function becomes

$$\ln \Xi_q = -\beta U_0 \ln e_g - \sum_i \ln\left(1-e_g^{-\beta \hbar \omega_i}\right) \approx -\beta 3Nu_0 \ln e_g, \qquad (54)$$

where in view of (28) there is

$$u_0 = \frac{\varphi_0}{3N} + \frac{3}{8}\hbar \omega_D < 0. \qquad (55)$$

In Eq. (55), the Debye model is also considered. Then in light of Eq.(49), we have

$$c_q = \Xi_q^{1-q} = e_g^{-\beta 3N(1-q)u_0} = e_g^{-\beta(1-\nu)u_0}. \qquad (56)$$

On the basis of (56), the generalized internal energy is written as

$$U = U_0 + 3Nk \frac{\pi^4}{5(\ln e_g)^4} \frac{T^4}{\theta_D^3} e_g^{\frac{4(v-1)u_0}{kT}}$$

$$\approx U_0 + 3Nk \frac{\pi^4}{5(\ln e_g)^4} \frac{T^4}{\theta_D^3} \left(1 + \frac{4(v-1)u_0}{kT}\right). \tag{57}$$

where only the first-order approximation of Tsallis factor is considered for the small nonextensive parameter (*v*-1) [45]. The generalized heat capacity is then given by

$$C_V = 3Nk \left( \frac{4\pi^4}{5(\ln e_g)^4} \frac{T^3}{\theta_D^3} + \frac{12\pi^4(v-1)}{5(\ln e_g)^3} \frac{T^2}{\theta_D^3} \frac{u_0}{k} \right). \tag{58}$$

It is interesting that in the above expression, there exists an additional term obeying the law of $T^2$. It is clearly that Eq.(58) and Eq.(58) both depend significantly on the nonextensive parameter ν and when we take $v \to 1$, they returns to the result based on the classical Debye model.

4.4 *An application of the generalized Debye model*

In this subsection, as an application we try to apply the generalized Debye model to study the heat capacity of complex solid through numerical method, and we mainly focus on the transition region of temperature, which can not be intuitively observed in the above analytical analyses.

For this aim, in view of (39) we consider the general expression of heat capacity, namely,

$$C_V = \frac{dU}{dT} = 3Nk \frac{3\ln(e_g)}{x^3} \int_0^x \frac{y^4}{(e_g^y - 1)^2} e_g^y dy, \tag{59}$$

where

$$y \approx \frac{\hbar\omega}{kT}, \quad x \approx \frac{\hbar\omega_D}{kT} = \frac{\theta_D}{T}. \tag{60}$$

Here, without loss of generality in physics, we have neglected the difference between the physical temperature $T_q$ and the Lagrange temperature *T*. In addition, for the convenience of analysis, we introduce the reduced heat capacity and the reduced temperature respectively as follows,

$$c_v = \frac{C_V}{3Nk}, \quad t = \frac{T}{\theta_D} = \frac{1}{x}. \tag{61}$$

The numerical analysis of the heat capacity based on the generalized Debye model for five values of parameter ν are given in Fig. 1, where the five special cases are represented by the short-dash line with ν=0.90, dot line withν=0.95, solid line with ν=1.00, long-dash line with ν=1.05 and dot-dash line with ν=1.10, respectively. It is clearly that the model curves depend sensitively on the value of the nonextensive parameter.

Furthermore, it should be noticeable that here the value of ν is assumed to be irrelevant to the solid temperature. The generalized Debye model might describe such a complex solid system where the non-nearest neighbor interaction between particles is not negligible, in which nonextensive statistical approach with the parameter ν may be applicable.

In more complex solid system, such as metal crystal where the long-range interaction between free electrons can not be ignored, the nonextensive parameter ν might be slightly dependent on the temperature of system in some range [7]. Mohammadzadeh et al gave an experimental data of the heat capacity of metal rubidium as a function of the temperature in Fig.9 of Ref.[47]. As an example, if we assume the nonextensive parameter ν to be related to the temperature $t$ by the relation,

$$\nu = \begin{cases} 1, & t > 50, \\ 0.94 + 0.042t, & t \leq 50, \end{cases} \quad (62)$$

then the numerical analysis of the heat capacity based on the generalized Debye model can be shown with the curve in Fig 2, marked by dot line. For comparison, the heat capacity based on the standard Debye model is also shown in Fig.2, marked by solid line. We can see that the profile of dot line based on the generalized Debye model is in excellent agreement with the experimental data of the heat capacity of metal rubidium in [47].

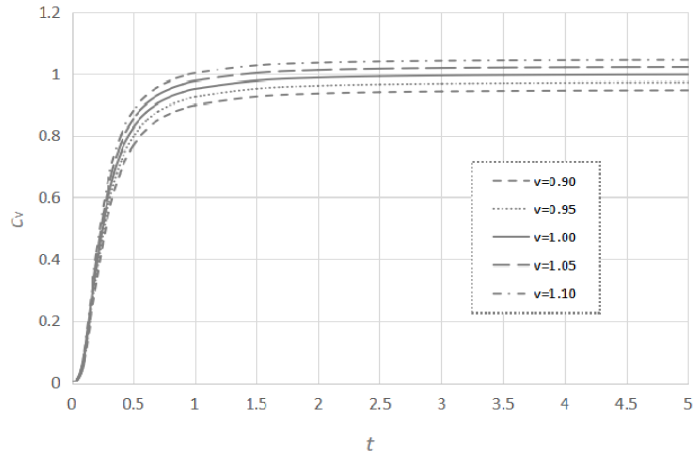

Fig 1. The heat capacity based on the generalized Debye model with five different nonextensive parameter.

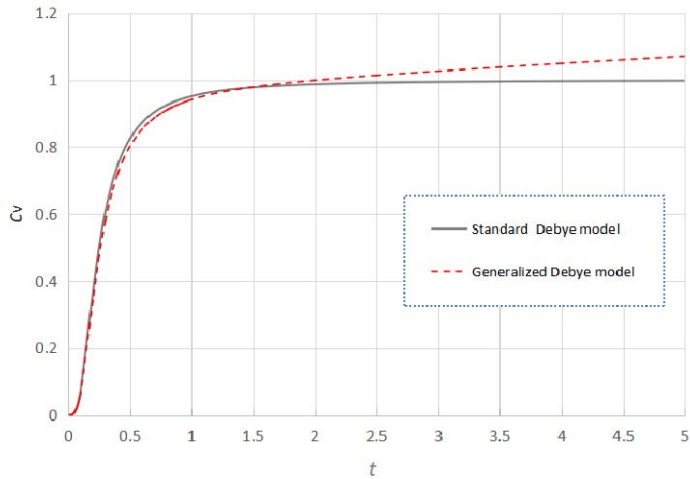

Fig 2. The heat capacity based on the generalized Debye model with the relation (62).

## 5 Property of metal electrons in the nonextensive quantum statistics

The electrons in a metal can move under the background of mean field of Coulomb long-range interactions which is equivalent to the short Coulomb shield effect. At the same time, the number density of electrons can be very high. So the nonextensive effect and the quantum effect should be both considered. This means that the electrons in such a metal should obey the Fermi distribution in the nonextensive quantum statistics. According to Eq.(22), the averaged occupation number at an energy level $l$ is described by the generalized Fermi distribution,

$$\bar{a}_l = \frac{g_l}{e_g^{(\varepsilon_l-\mu)/kT_q}+1}, \tag{63}$$

where $g_l$ is the degeneracy of electrons. The state density of electrons [46] is

$$g(\varepsilon)d\varepsilon = \frac{4\pi V}{(2\pi\hbar)^3}(2m)^{3/2}\sqrt{\varepsilon}d\varepsilon, \tag{64}$$

where $m$ is mass of the electron and $V$ is volume of the electrons. According to Eqs.(14) and (15), it can be proved that the total (Lagrange) electron number reads

$$N = \frac{4\pi V}{(2\pi\hbar)^3}(2m)^{3/2}\int_0^\infty \frac{\sqrt{\varepsilon}d\varepsilon}{e_g^{(\varepsilon-\mu)/kT_q}+1}. \tag{65}$$

Similarly, the internal energy of the electrons is written as

$$U = \frac{4\pi V}{(2\pi\hbar)^3}(2m)^{3/2}\int_0^\infty \frac{\sqrt{\varepsilon^3}d\varepsilon}{e_g^{(\varepsilon-\mu)/kT_q}+1}. \tag{66}$$

Generally, the chemical potential of the electrons is a function of temperature. Now let us consider the limit case of zero physical temperature $T_q=0$, at which the averaged occupation number at each energy quantum state should be

$$\frac{\bar{a}_l}{g_l} = \begin{cases} 1, & \varepsilon < \mu(0) \\ 0, & \varepsilon > \mu(0) \end{cases}, \tag{67}$$

where $\mu(0)$ is the chemical potential at zero point of the physical temperature. Here, we mention that, according to Eq.(38), the zero point of the physical temperature is also the zero point of the Lagrange temperature. According to Pauli exclusion principle, the quantity $\mu(0)$ is also the maximum energy of the electrons, called as Fermi energy [46]. In view of Eqs.(65) and (66), at the zero point of the (physical or Lagrange) temperature, we obtain the electrons number:

$$N = \frac{4\pi V}{(2\pi\hbar)^3}(2m)^{3/2}\frac{2}{3}[\mu(0)]^{3/2} \equiv \frac{2}{3}A[\mu(0)]^{3/2}, \tag{68}$$

and the internal energy:

$$U(0) = \frac{4\pi V}{(2\pi\hbar)^3}(2m)^{3/2}\int_0^{\mu(0,n)}\sqrt{\varepsilon}d\varepsilon = \frac{3}{5}N\mu(0). \tag{69}$$

In order to calculate the chemical potential and internal energy of electrons at nonzero temperature, we need to evaluate the following integral,

$$I \equiv \int_0^\infty \frac{F(\varepsilon)}{e_g^{(\varepsilon-\mu)/kT_q}+1}d\varepsilon, \tag{70}$$

where $F$ is a specified function. Defining $x=(\varepsilon-\mu)/kT_q$, we get that

$$I = kT_q \int_{-\mu/kT_q}^{\infty} \frac{F(\mu + kT_q x)}{e_g^{\ x} + 1} dx$$

$$= kT_q \int_{0}^{\mu/kT_q} \frac{F(\mu - kT_q x)}{e_g^{\ -x} + 1} dx + kT_q \int_{0}^{\infty} \frac{F(\mu + kT_q x)}{e_g^{\ x} + 1} dx$$

$$\approx \int_{0}^{\mu} F(\varepsilon) d\varepsilon + kT_q \int_{0}^{\infty} \frac{F(\mu + kT_q x) - F(\mu - kT_q x)}{e_g^{\ x} + 1} dx . \qquad (71)$$

On the right-hand side of (71), the main part in the integrand of the second term is around $x=0$, so we should have

$$I \approx \int_{0}^{\mu} F(\varepsilon) d\varepsilon + 2(kT_q)^2 F'(\mu) \int_{0}^{\infty} \frac{x}{e_g^{\ x} + 1} dx$$

$$= \int_{0}^{\mu} F(\varepsilon) d\varepsilon + \frac{\pi^2}{6} \left(\frac{kT_q}{\ln e_g}\right)^2 F'(\mu) . \qquad (72)$$

Substituting $F(\varepsilon) = \varepsilon^{1/2}$ and $F(\varepsilon) = \varepsilon^{3/2}$ into Eq.(72), we obtain the particle number and the internal energy of the electrons, given respectively by

$$N = \frac{2}{3} A \mu^{3/2} \left( 1 + \frac{\pi^2}{8} \left( \frac{kT_q}{\mu \ln e_g} \right)^2 \right), \qquad (73)$$

$$U = \frac{2}{5} A \mu^{5/2} \left( 1 + \frac{5\pi^2}{8} \left( \frac{kT_q}{\mu \ln e_g} \right)^2 \right). \qquad (74)$$

It can be seen that at a room temperature, the thermal motion energy $kT$ (the temperature $T$ is identical to the thermodynamic temperature in the classical statistics [41]) is very small relative to the chemical $\mu$. Therefore, the correction effect caused by the Tsallis factor is very small and can be ignored, so indicating that $c_q=1$. Thus, the physical temperature can be directly replaced by the Lagrange temperature in Eqs.(73) and (74). Then, combining Eq. (68) with Eq.(73), we obtain the generalized chemical potential of the electrons,

$$\mu \approx \mu(0) \left( 1 - \frac{\pi^2}{12} \left( \frac{kT}{\mu(0) \ln e_g} \right)^2 \right). \qquad (75)$$

Substituting Eq. (68) and Eq.(75) into Eq.(74), the generalized internal energy is written as

$$U \approx \frac{3}{5} N \mu(0) \left( 1 + \frac{5\pi^2}{12} \left( \frac{kT}{\mu(0) \ln e_g} \right)^2 \right). \qquad (76)$$

Therefore, the generalized heat capacity of the electrons in the metal in the nonextensive quantum statistics is

$$C_V = Nk \frac{\pi^2}{2(\ln e_g)^2} \frac{kT}{\mu(0)}. \tag{77}$$

Clearly, Eqs.(76) and (77) both depend significantly on the nonextensive parameter $v$ and when we take $v \to 1$, they return to the results of that in the classical quantum statistics.

## 6 Conclusions

In this work, we have studied the grand canonical distribution function in the nonextensive quantum statistics. With a new parameter transformation, we deal with the factorization problem in the power-law distribution and derive the generalized Boson distribution and Fermi distribution in the nonextensive quantum statistics, given in Eq.(22) and Eq.(23) respectively. Such a parameter transformation also introduces a generalized Nature Index in the form of (31) and thus we produce a generalized Debye model (41) in the nonextensive quantum statistics. These new expressions all depend significantly on the nonextensive parameter $v$ and when we take $v \to 1$, they all recover to the forms in the classical quantum statistics.

By using the generalized Debye model in the nonextensive quantum statistics, we derive the expressions of the internal energies and capacities of the solid in the limit case of high temperature and low temperature, respectively, given by Eqs. (50) and (51), and Eqs.(57) and (58), which significantly depend on the nonextensive parameter $v$. When we take $v \to 1$, they all recover to the results based on the standard Debye model in the classical quantum statistics. Some numerical results of the heat capacity based on the generalized Debye model are shown in Fig 1 and Fig 2. The nonextensivity in complex solid system might be coming from the non-nearest neighbor interaction between atoms or /and the long-range Coulomb interaction between metal electrons.

Further, by using the nonextensive quantum statistics we study the chemical potential, internal energy and heat capacity of the metal electrons with the generalized Fermi distribution, where the Coulomb long-range interactions may play a role. We have derived the expressions of the chemical potential, the internal energy and the heat capacity of electrons, given by Eqs.(75), (76) and (77), respectively, in which we show that the thermodynamic quantities depend significantly on the nonextensive parameter and when we take the limit $v \to 1$, they all recover to the results based on the Fermi distribution in the classical quantum statistics.

Acknowledgements

This work is supported by the National Natural Science Foundation of China under Grant No. 11405092, and also by National Natural Science Foundation of China under Grant No. 11775156.